\documentclass[twocolumn,floatfix,altaffilletter,superscriptaddress,preprintnumbers,
               tightenlines,showpacs,showkeys,nofootinbib]{revtex4-1}

\usepackage[dvips]{graphicx}

\usepackage{siunitx}
\usepackage{array,amsmath,amsthm,feynmf,mathtools}
\usepackage{mathtools}
\usepackage{epsfig}
\usepackage{graphicx}
\usepackage{color}
\usepackage{slashed}
\usepackage{amssymb}

\makeatletter
\let\saved@underbrace\underbrace
\renewcommand*\underbrace[1]{\@ifnextchar_{\ub@with{#1}}{\ub@without{#1}}}
\def\ub@with#1_#2{\mathpalette\underbrace@i{{#1}{_{#2}}}}
\newcommand*\ub@without[1]{\mathpalette\underbrace@i{{#1}{}}}
\newcommand*\underbrace@i[2]{\underbrace@ii#1#2}
\newcommand*\underbrace@ii[3]{\saved@underbrace{#1#2}#3}
\makeatother

\begin{document}

\title{Loop Dominated Signals from Neutrino Portal Dark Matter}
\preprint{CERN-TH-2019-203}
\author{Hiren H. Patel}
\email{hpatel6@ucsc.edu}
\affiliation{Santa Cruz Institute for Particle Physics, University of California, Santa Cruz, CA 95064, USA}
\affiliation{Department of Physics, 1156 High St., University of California Santa Cruz, Santa Cruz, CA 95064, USA}

\author{Stefano Profumo}
\email{profumo@ucsc.edu}
\affiliation{Santa Cruz Institute for Particle Physics, University of California, Santa Cruz, CA 95064, USA}
\affiliation{Department of Physics, 1156 High St., University of California Santa Cruz, Santa Cruz, CA 95064, USA}

\author{Bibhushan Shakya }
\email{bibhushan.shakya@cern.ch}
\affiliation{CERN, Theoretical Physics Department, CERN, 1211 Geneva 23, Switzerland}

\begin{abstract}
%
We study scenarios where loop processes give the dominant contributions to dark matter decay or annihilation despite the presence of tree level channels. We illustrate this possibility in a specific model where dark matter is part of a hidden sector that communicates with the Standard Model sector via a heavy neutrino portal. We explain the underpinning rationale for how loop processes mediated by the portal neutrinos can parametrically dominate over tree level decay channels, and demonstrate that this qualitatively changes the indirect detection signals in positrons, neutrinos, and gamma rays. 
\end{abstract}

\maketitle

\section{Motivation}

The microscopic nature of dark matter remains one of the most pressing questions in particle physics. Indirect detection---the search for visible signatures of dark matter decay or annihilation at terrestrial or space-based experiments---is one of the leading programs to unravel this mystery. In this paper, we study the possibly dominant role of loop diagrams to dark matter annihilation/decay processes and subsequent indirect detection signatures in frameworks where dark matter is part of a secluded or hidden sector that couples weakly to the Standard Model (SM) sector via a neutrino portal \cite{Lindner:2010rr,Cherry:2014xra,Roland:2014vba,Shakya:2015xnx,Gonzalez-Macias:2016vxy,Escudero:2016tzx,Escudero:2016ksa,Schmaltz:2017oov,Batell:2017cmf,Batell:2017rol,Shakya:2018qzg,Blennow:2019fhy}. In the context of dark matter annihilation or decay, loop diagrams generally become important when tree level processes are forbidden for some reason, \textit{e.g.} for line signals in gamma rays, but are otherwise only expected to produce subleading corrections. However, the dominance of loop processes when tree level channels in the same final states are open is more subtle and interesting, and important for phenomenology.

From theoretical considerations, a neutrino portal to a dark sector is known to be one of only a few ways to connect visible and hidden sectors via renormalizable interactions, and  the existence of dark matter in such setups has been extensively studied in several earlier works \cite{Lindner:2010rr,Cherry:2014xra,Roland:2014vba,Shakya:2015xnx,Gonzalez-Macias:2016vxy,Escudero:2016tzx,Escudero:2016ksa,Schmaltz:2017oov,Batell:2017cmf,Batell:2017rol,Shakya:2018qzg,Blennow:2019fhy}. Such frameworks are particularly motivated in light of models of neutrino mass generation, which requires physics beyond the Standard Model, including SM-singlet new states (sterile neutrinos), which can act as portals to hidden sectors. Neutrino-rich indirect detection signatures in such models have been extensively studied in the literature \cite{Garcia-Cely:2017oco,ElAisati:2017ppn,Campo:2017nwh,Chianese:2018ijk,Blennow:2019fhy,Dekker:2019gpe,Heeck:2019guh}. From the point of view of phenomenology, indirect detection of dark matter in neutrino-rich final states has recently garnered tremendous interest in the community, driven by sensitive instruments such as IceCube \cite{Aartsen:2019swn}, Super-Kamiokande \cite{Frankiewicz:2015zma}, and ANTARES \cite{Tonnis:2019hyr}, and have also been fuelled by anomalous high energy neutrino events at IceCube \cite{Aartsen:2013jdh,Aartsen:2014gkd}, which can be interpreted as hints of decaying dark matter (see eg. \cite{Cohen:2016uyg,Roland:2015yoa} and references therein).  

Motivated by such considerations, the main purpose of this paper is to demonstrate that loop processes, often ignored, can dominate dark matter decay or annihilation in realistic scenarios of neutrino portal dark matter. For concreteness, we describe this effect in a specific model of decaying hidden sector $Z'$ dark matter (Section \ref{sec:framework}), where the loop process is manifestly finite and can be calculated explicitly (Section \ref{sec:calculation}). However, we point out that the dominance of loop processes over tree level processes is more general and can occur in several other frameworks (Section \ref{sec:others}). We also study the implications of this effect on the spectra of SM particles (neutrinos, positrons, and gamma rays) from dark matter, relevant for indirect detection (Section \ref{sec:pheno}). 

\section{Framework}
\label{sec:framework}

We base our discussions on a model of hidden sector $Z'$ dark matter, which illustrates the main ideas of this paper in the most straightforward manner. The model consists of sterile neutrinos in an extended, secluded sector, similar in spirit to the frameworks studied in \cite{Shakya:2018qzg,Roland:2014vba,Roland:2015yoa,Roland:2016gli,Chacko:2016hvu}. We consider a dark gauged $\text{U}(1)'$ sector, with gauge coupling $g'$ and a corresponding gauge boson $Z'$, and three new categories of fields: a fermion $\nu'$ and a singlet scalar $S$ with $\text{U}(1)'$ charges $+1,-1$ respectively, as well as completely singlet sterile neutrinos $N_i$, which carry no $\text{U}(1)'$ or SM charges. Note that $\nu'$ and $S$ can be thought of as hidden sectors analogs of the SM neutrinos and Higgs, which can be combined into a gauge singlet and therefore couple to $N_i$ via a renormalizable Dirac mass term. The Lagrangian for this model is
\begin{multline}\label{eq:modellag}
\mathcal{L}=|D_\mu S|^2 - V(H,S) - \frac{1}{4}F'_{\mu\nu}F'^{\mu\nu}  + \nu'^\dag i \bar{\sigma}^\mu D_\mu \nu' + N_i^\dag i \bar{\sigma}^\mu \partial_\mu N_i\\ - \frac{1}{2} (M_A N_A^\dag N_A^\dag + \theta_N M_{AB} N_A^\dag N_B^\dag+M_B N_B^\dag N_B^\dag +\text{c.c.}) \\
- (y' S\nu'^\dag N_A^\dag + \text{c.c.}) - (y_\nu \tilde{H}L^\dag N_B^\dag + \text{c.c.})\,,
\end{multline}
where $D_\mu = \partial_\mu + i g' Z'_\mu$. We assume that $H$ and $S$ acquire vacuum expectation values (vev) $v$ and $x$ respectively, spontaneously breaking the electroweak and $\text{U(1)}'$ symmetries.

We consider two sets of heavy singlet sterile neutrinos\,\footnote{The exact number of heavy sterile neutrinos is irrelevant for our discussions, hence we leave it unspecified.}, $N_A$ and $N_B$, that couple dominantly to the hidden and visible sector respectively. We will consider the singlet neutrino mass scale $M_N\approx M_A \approx M_B \approx M_{AB}$ to be heavier than all other scales in the theory. The sterile neutrinos $N_B$ give rise to neutrino masses $m_\nu\approx y_\nu^2 v^2/M_N$ for the SM neutrinos via the well known type-I seesaw mechanism \cite{Minkowski:1977sc}. Furthermore, this heavy sterile neutrino sector acts as the portal between the visible and secluded sectors via the mass cross-term $\theta_N M_{AB}$, where $\theta_N$ has been introduced to control the size of the mixing between the two sectors. 

Spontaneous $\text{U(1)}'$ breaking gives the $Z'$ a mass $m_{Z'} = g'x/2$, and the SM-singlet neutrinos $\nu', N_A$, $N_B$ mix to form mass eigenstates $N_1,N_2, N_3$ with masses $M_1,M_2, M_3$ respectively. For $M_N\gg y'x$, we also have a seesaw effect in the hidden sector, resulting in $M_1\approx y'^2 x^2/M_N$ and $M_2, M_3\approx M_N$. Upon electroweak symmetry breaking (we will treat it as a perturbative effect), all three of these mass eigenstates inherit small mixings with the SM neutrinos via the Dirac mass terms.

We are interested in the parameter space where $Z'$ is the lightest hidden sector particle and therefore the dark matter. We thus focus on the hierarchy $m_Z' < M_1$, so that the $Z'$ cannot decay into the $N_1$ states at tree level\,\footnote{Scenarios where sterile neutrinos are light enough to be produced directly in dark matter annihilation or decay can also produce interesting dark matter signatures (see e.g. \cite{Escudero:2016tzx,Roland:2015yoa,Capozzi:2017auw,Folgado:2018qlv,Gori:2018lem}), but we do not consider such scenarios in this paper.}. It can decay into SM neutrinos via neutrino mixing between the two sectors; however, the lifetime for this process can be sufficiently long that $Z'$ remains a viable dark matter candidate. The free parameters in this setup are $\theta_N,g',x,y_\nu,y_\nu',$ and $M_N$. One can trade the latter four parameters for the three neutrino mass scales, $m_\nu, M_1$, and $M_2$, and the dark matter mass $m_{Z'}$. The remaining free parameters $g'$ and $\theta_N$ can then be used appropriately to set the dark matter relic abundance and lifetime.

We neglect the kinetic mixing term $\frac{\epsilon}{2}F^{\mu\nu}Z'_{\mu\nu}$ between the hypercharge and dark $Z'$ gauge boson. This mixing, even if absent at tree level, is generally generated by loop effects in the presence of heavy particles that couple to both gauge fields; however, in the model above, such mixing is only generated at three loops (involving the secluded fermion $\nu'$, the heavy mediators $N$, and the SM neutrinos $\nu$) and is therefore expected to be negligible. Likewise, we also assume that the renormalizable Higgs portal coupling $S^2h^2$ is negligible. Finally, we assume that additional heavy matter content is present to ensure the $\text{U(1)}'$ dark current remains anomaly free as needed without affecting the decay processes we consider; we will comment further on this later. 

\section{Evaluation of Loop Processes}
\label{sec:calculation}

\begin{figure}[t]
\includegraphics[width=0.25\columnwidth]{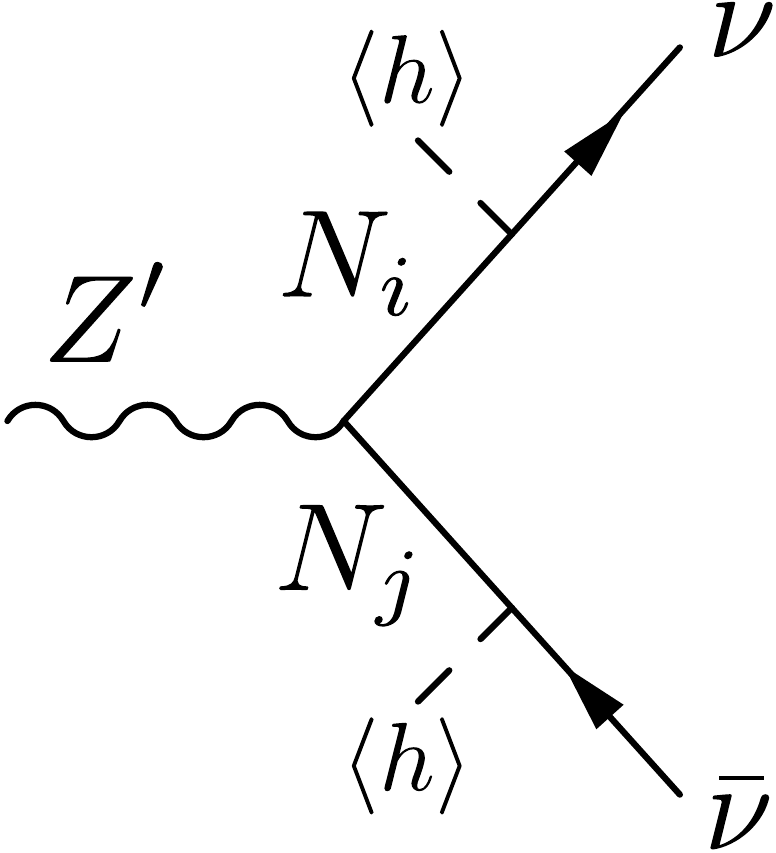}\enspace
~~\includegraphics[width=0.25\columnwidth]{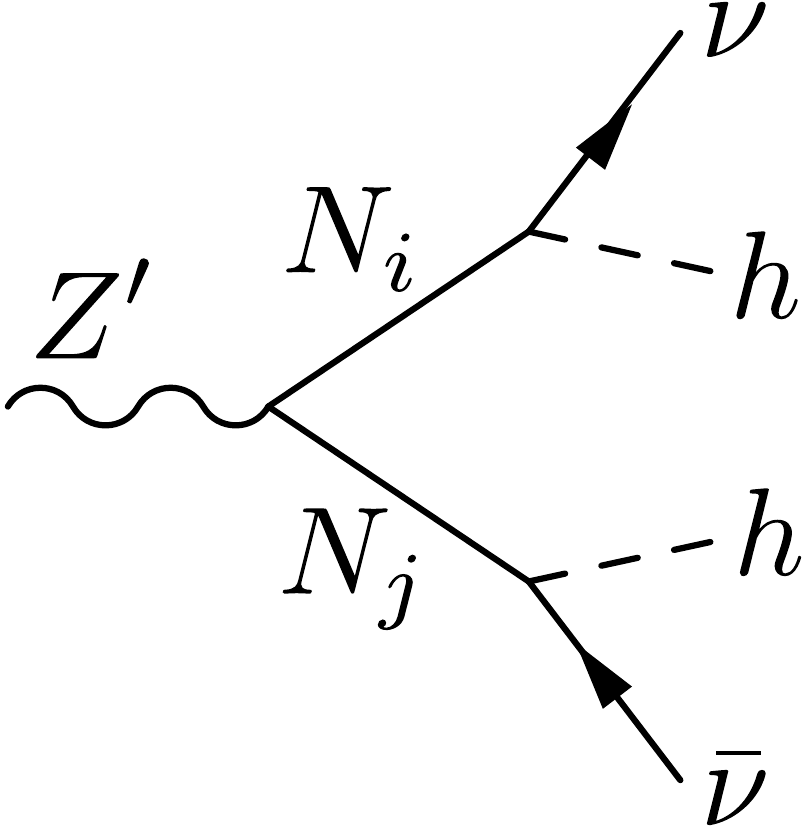}\enspace
~~~\includegraphics[width=0.25\columnwidth]{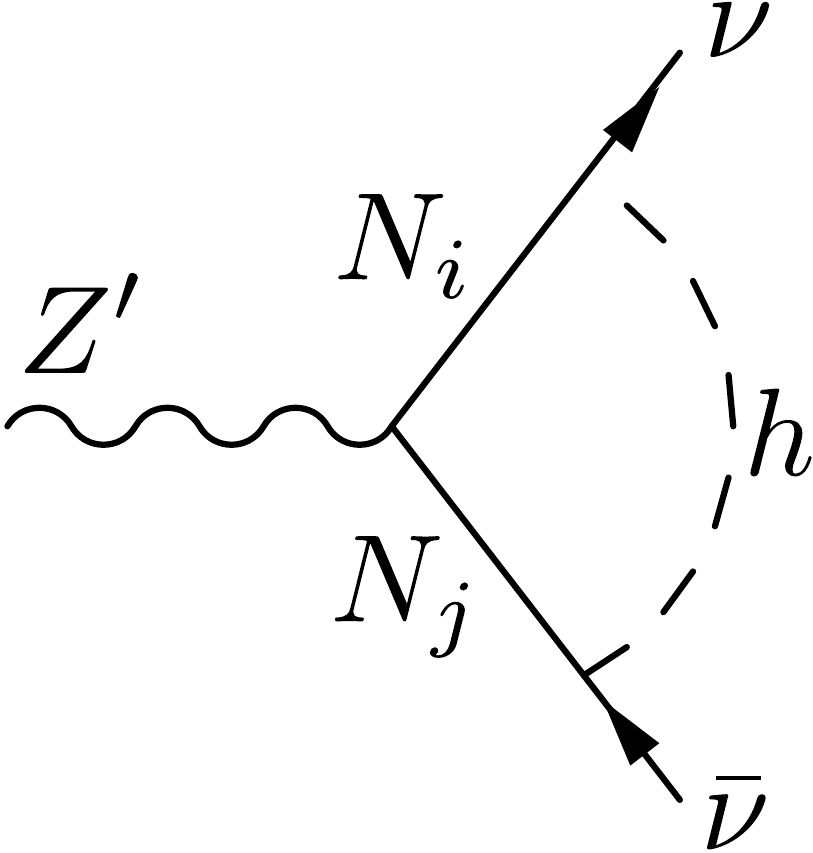}
\caption{\label{fig:diagrams} Dark matter decay modes: tree-level two body, tree-level four body, and one loop. This list is not exhaustive; we only show a representative set of decay modes (see text).}
\end{figure}

The leading tree and loop diagrams for dark matter decay in this model are shown in Figure\,\ref{fig:diagrams}. The leading tree level decay, represented by the first diagram, is into two neutrinos:
\begin{equation}
\Gamma_{2t}=\frac{m_{Z'}g'^2 \theta_N^4 y_\nu^4}{48\pi}\frac{v^4}{M_1^2 M_2^2}=\frac{m_{Z'}g'^2 \theta_N^4}{48\pi}\frac{m_{\nu}^2}{M_1^2},
\end{equation}
where, in the second step, we have used the seesaw relation $m_\nu=y_\nu^2 v^2/M_N$. Note that this decay width is suppressed by the active-sterile mixing angle, represented by Higgs vev insertions in the diagram: this is particularly clear from the second expression above, where $\theta_N \sqrt{m_\nu/M_1}$ is the effective mixing angle between $\nu$ and $N_1\approx \nu'$, which is the fermion state that couples directly to $Z'$ with gauge coupling strength $g'$. 

If the dark matter is sufficiently heavy, additional three and four body decay channels that involve SM Higgs and gauge bosons become kinematically accessible. These diagrams can be understood as replacing the Higgs vevs that gives rise to the SM-singlet neutrino mixing with the emission of physical states, as shown schematically in Figure\,\ref{fig:diagrams}; due to the SU(2) nature of the SM neutrinos, there are additional diagrams that involve charged leptons and gauge bosons. The decay widths into these multi-body final states can be found, e.g., in \cite{Cohen:2016uyg}. Replacing a Higgs vev on one of the neutrino legs with the emission of a physical particle gives rise to three body decay channels $\nu\bar{\nu} h,\,\nu\bar{\nu} Z\,,\nu \bar{l} W$ with decay widths \cite{Cohen:2016uyg}
 \begin{equation}
\Gamma_{3t}\approx\frac{m_{Z'}^2}{768\pi^2 v^2}\Gamma_{2t}.
\label{eq:threebody}
\end{equation}
Here $l$ refers to a charged lepton, whose flavor depends on the flavor of the SM neutrino that couples to the portal states. Likewise, the four body decay channels, obtained by replacing vevs on both neutrino legs with physical particle emissions, include the final states $\nu\bar{\nu} h h,\,\nu\bar{\nu} Z Z, \nu\bar{\nu}Z h,\,\nu \bar{l} h W,\,\nu \bar{l} Z W, l\bar{l}WW$; the decay widths for these processes scale (up to some $\mathcal{O}(1)$ factors) as \cite{Cohen:2016uyg}
\begin{equation}
\Gamma_{4t}\approx\frac{m_{Z'}^2}{24\pi^2 v^2}\Gamma_{3t}.
\label{eq:fourbody}
\end{equation}

Note that the expression of four and three body decay widths in terms of three and two body widths is very intuitive: one incurs additional phase space suppression due to the emission of an additional particle, but gains a factor of $m_Z'^2/v^2$ because the Higgs vev insertion gets replaced by the energy scale of the process, which is the dark matter mass. Thus, for sufficiently heavy dark matter $m_{Z'}\gg v$, the four body process can dominate: since electroweak symmetry breaking is effectively a small perturbation in this limit, the emission of a physical Higgs boson, which can proceed in the limit of unbroken electroweak symmetry, is preferred.  

We now turn to the evaluation of the loop processes shown in  Fig.\,\ref{fig:diagrams}, where the Higgs vev insertions are replaced by a Higgs propagator, giving rise to a one loop contribution to $Z'\to\nu\bar{\nu}$. Loop diagrams of this form are generally divergent; however, this diagram is manifestly finite in our framework by construction due to the absence of tree level couplings of SM neutrinos or gauge bosons to the $Z'$. This contribution can therefore be unambiguously evaluated. We calculate the full two-body decay width for $Z'\to \nu\bar{\nu}$, including the loop correction, under the approximation $m_h,\, m_{Z'} \ll M_1 \ll M_N$, to be
\begin{equation}
\Gamma_{2} = \frac{m_{Z'}g'^2 \theta_N^4 y_\nu^4}{48\pi}\Big|\frac{v^2}{M_1 M_2} + \frac{1}{32\pi^2}\frac{M_1}{M_2}\big(\ln\frac{M_2^2}{M_1^2} + 1\big)\Big|^2\,.
\label{fig:loopfull}
\end{equation}
We find that the naive log-divergence of the loop process is rendered finite upon summing over the various sterile neutrino propagator combinations in the loop, leaving behind the finite logarithm $\ln(M_2^2/M_1^2)$. The factor of $M_1/M_2$ in front of the logarithm represents the mixing angle between the $\nu'$ and $N_A$ states. In the limit where the tree level contribution can be neglected, the width $Z'\to\nu\bar{\nu}$ due to the loop process is 
\begin{equation}
\Gamma_{2} = \frac{m_{Z'}g'^2 \theta_N^4 }{48\pi (32\pi^2)^2} \frac{m_\nu^2 M_1^2}{v^4}\big(\ln\frac{M_2^2}{M_1^2} + 1\big)^2.
\label{eq:loop}
\end{equation}

Again, due to the SU(2) nature of the SM neutrinos, there are analogous loop processes for decays into other SM final states that scale in the same manner. Replacing the neutral SU(2) states with charged SU(2) states results in a W-loop induced decay into charged leptons $Z'\to l^+l^-$ with the same amplitude as $Z'\to \nu\bar{\nu}$ above (note that the analogous tree level decay process into charged leptons does not exist, since the charged components of the Higgs field do not obtain vevs). Likewise, ``flipping" the external legs and closing the loop with fermions instead of bosons gives rise to $Z'\to Zh, W^+W^-$ at one loop. In the limit $m_h, m_Z \ll m_{Z'} \ll M_1 \ll M_N$, we evaluate these widths to be
\begin{gather}
\begin{aligned}
\label{eq:loop2}
\Gamma_{Z'\to Zh} &= \frac{m_{Z'}g'^2 \theta_N^4 }{64\pi (32\pi^2)^2} \frac{m_\nu^2 M_1^2}{v^4}\big(1-\ln\frac{M_2^2}{M_1^2} \big)^2\,,\\
\Gamma_{Z'\to WW}& = 2\,\Gamma_{Z'\to Zh}\,.
\end{aligned}
\end{gather}
These are parametrically the same as the loop-induced widths to fermions above, up to $\mathcal{O}(1)$ factors. Due to Bose symmetry, $\Gamma_{Z'\to hh}$ vanishes, while $\Gamma_{Z'\to ZZ}$ is suppressed \cite{Keung:2008ve} relative to the above widths by a factor $\sim 12 m_Z^2/m_Z'^2$ according to our calculations, which renders it negligible for dark matter at the TeV scale or higher.

It is now illustrative to compare the loop dominated decay width in Eq.\,(\ref{eq:loop}) to the leading two body and four body tree level decay widths:
\begin{equation}
{\Gamma_{2t}}:{\Gamma_{4t}}:{\Gamma_{2}}\approx v^4:\frac{M_{Z'}^4}{18 (32\pi^2)^2}:\frac{M_1^4}{(32\pi^2)^2}\, \big(\ln\frac{M_2^2}{M_1^2} + 1\big)^2,
\end{equation}
From this comparison, we see that the loop processes can dominate if $M_1> m_{Z'}, 10\, v$ (recall that $M_1>m_{Z'}$ is an underlying assumption of our model for $Z'$ to be the lightest particle in the hidden sector). The origin of this domination is also clear from the above discussions. The two body decays require mixing between active and sterile neutrinos on both neutrino legs, and are therefore suppressed by $v^4$ from the associated Higgs vev insertions. The four-body decays do not require electroweak symmetry to be broken and therefore avoid this suppression, depending instead on the relevant energy scale of the process, $m_{Z'}$. The loop diagram (which can also proceed without electroweak symmetry breaking) avoids even this (milder) suppression, as the relevant energy scale is instead the sterile neutrino mass $M_1$. 

Finally, we also note the existence of two-loop diagrams (obtained by closing the singlet Higgs $S$ loop on the ``hidden sector" side, together with the SM Higgs loop)  that can contribute to dark matter decay in the above framework. Relative to the one loop diagram, this process incurs additional loop suppression but could evade the $M_1^2/M_2^2$ suppression in Eq.\,\ref{fig:loopfull} (recall that this represents a mixing angle between $\nu'$ and $N_A$), which can be significant in the regime $M_1\ll M_2$. For simplicity, in this paper we will restrict ourselves to the regime where the ratio $M_1/M_2$ is sufficiently large that the two-loop contribution is subdominant and can be ignored.

\section{Other Scenarios}
\label{sec:others}

In this section, we present a broader discussion of the importance of the loop process in other scenarios, shedding further light on the conditions necessary for the loop process to dominate dark matter phenomenology. 

A particularly well motivated dark matter candidate that couples preferentially to neutrinos is the Majoron $J$ \cite{Gelmini:1980re,Boulebnane:2017fxw,Pilaftsis:1993af,Rothstein:1992rh,Berezinsky:1993fm,Frigerio:2011in,Lattanzi:2007ux,Bazzocchi:2008fh,Lattanzi:2013uza,Queiroz:2014yna}, the Goldstone boson associated with spontaneously broken lepton number. While one loop contributions to the decay $J\to\nu\nu$ exist, they are always subdominant to the tree level processes, in contrast to the above discussions. This discrepancy can be understood by following the flow of lepton number and hypercharge: the Majoron carries lepton number $+2$ but no hypercharge; on the other hand, the final state $\nu\nu$, enforced by lepton number conservation, carries two units of hypercharge. This must therefore be balanced by two Higgs vev insertions in the decay process. This is already present a priori in the two body decay process in the form of active-sterile mixing, but the loop diagram requires explicit vev insertions, which suppresses it and keeps it subdominant to the two-body tree level diagram. Therefore, loop processes of the kind discussed above only provide subleading corrections for Majoron dark matter. 

The decay of scalar dark matter into neutrinos through the heavy neutrino portal, meanwhile, is helicity suppressed and must necessarily pick up factors of neutrino masses $m_\nu$ (or equivalently, Higgs vevs), hence the loop process cannot overcome the $\sim v^4$ suppression factor present in the tree-level diagram. One can consider other dark matter scenarios that avoid this helicity suppression, e.g. annihilation of a dark matter fermion $\chi$ into neutrinos mediated by dimension-6 current$\times$current operators $\chi^\dag\bar\sigma^\mu\chi N^\dag\bar\sigma_\mu N$: The amplitude for the loop mediated decay to neutrinos in this case, however, is UV divergent, and an unambiguous prediction of its size independent of the details of the UV physics is not possible. Several other models, including a more naive implementation of a vector dark matter model, also suffer from this UV sensitivity.

Nevertheless, there exist other neutrino portal dark matter scenarios where the loop contribution is finite as well as dominant. If dark matter is a hidden sector fermion that annihilates via a heavy $Z'$ mediator into neutrinos, the above discussions are directly applicable, and dark matter annihilations can be dominated by loop processes. Likewise, loop dominance can also feature in supersymmetric theories: if a hidden sector gaugino is the lightest supersymmetric particle (LSP) and dark matter that primarily annihilates into SM neutrinos via exchange of a heavy sterile sneutrino in the $t$-channel, the loop-induced annihilation process obtained by extending this tree level diagram with a Higgs loop, analogous to our discussions in the previous sections, would also be finite as well as dominant.  

In such extended frameworks, it should be kept in mind that there might be additional loop processes contributing to dark matter annihilation or decay, involving additional particles required, for instance, for anomaly cancellation. For instance, if the underlying theory of the $Z'$ dark matter model discussed in Section \ref{sec:framework} is supersymmetric, one gets Higgsino-sneutrino loops that are supersymmetric counterparts to the loop diagrams that were considered. Such loop contributions are parametrically of the same form as those calculated above and can cause $\mathcal{O}(1)$ modifications of the dark matter annihilation or decay rate.  

Finally, it is worth pointing out that this loop dominated behavior is not confined to neutrino portal models but can be realized more broadly in any framework where leading tree level channels incur some form of suppression (analogous to the active-sterile mixing angle suppression in neutrino portal models) that can be lifted by considering loop processes.

\section{Dark Matter Phenomenology}
\label{sec:pheno}

We now turn to a discussion of the implications of loop dominance for dark matter phenomenology. Effects on dark matter production mechanism and lifetime, while likely significant, are model-specific aspects and therefore of limited applicability, hence we only discuss these briefly within our $Z'$ dark matter model. The effects on the annihilation or decay signatures that would be observed at indirect detection instruments, on the other hand, are model-independent and robust (in the sense that they hold more broadly for a greater class of neutrino portal models where the loop process dominates); thus we study these aspects in greater detail. 

\subsection{Dark Matter Parameter Space}
In the $Z'$ dark matter model from Section \ref{sec:framework}, if the mixing between the two sectors $\theta_N$ is small, the secluded sector does not thermalize with the SM thermal bath and is instead populated by freeze-in processes. The annihilation processes $\nu h\to \nu' S$, mediated by $N_{A,B}$ in the s-channel, produces small amounts of the secluded sector particles $\nu', S$. While the $\nu'$ tends to decay primarily into the visible sector via $\nu'\to \nu Z$, the scalar S decays primarily as $h'\to Z'Z'$ if $g'$ is larger than $y'x/M_2$ (which controls the other available decay channel $h'\to\nu'\nu'$), producing a small dark matter abundance. Since $\nu h\to \nu' S$ is a dimension 5 operator, the subsequent dark matter abundance depends on the reheat temperature $T_{RH}$, the highest temperature attained by the early radiation-dominated Universe; this abundance can be estimated as \cite{Hall:2009bx,Elahi:2014fsa}
\begin{equation}
Y_{Z'}\sim10^{-6}y_\nu^2 y'^2\theta_N^2\frac{M_{Pl} T_{RH}}{M_2^2}.
\end{equation}
Substituting the neutrino masses $m_\nu, M_1$ for the Yukawa couplings and plugging in the known values of $m_\nu, v,$ and $M_{Pl}$ (the Planck mass), we obtain the following relation in order to achieve the correct dark matter relic density:
\begin{equation}
\left(\frac{m_{Z'}}{\text{TeV}}\right)\left(\frac{\text{TeV}}{M_{1}}\right)\left(\frac{10\,\text{TeV}}{T_{RH}}\right)\sim\left(\frac{g'\,\theta_N}{10^{-5}}\right)^2
\end{equation}
The dark matter lifetime, on the other hand, is controlled by the leading (loop level) decay widths, which scale as $\Gamma\sim g'^2\theta_N^4$. With appropriate choices of $\theta_N, g',$ and $T_{RH}$, we can therefore achieve both the correct dark matter relic density as well as lifetimes that are interesting for indirect detection signals. As illustrative numbers, $g'\sim10^{-3},\theta_N\sim10^{-7}, m_{Z'}\sim$ TeV, $M_1\sim 100$ TeV, and $T_{RH}\sim10^7$ TeV lead to a consistent cosmology with the correct dark matter relic abundance, loop-dominated decays, and dark matter lifetime $\sim10^{28}$s. 

\subsection{Indirect Detection Signatures}

\begin{figure}[h!]
\includegraphics[width=0.85\columnwidth]{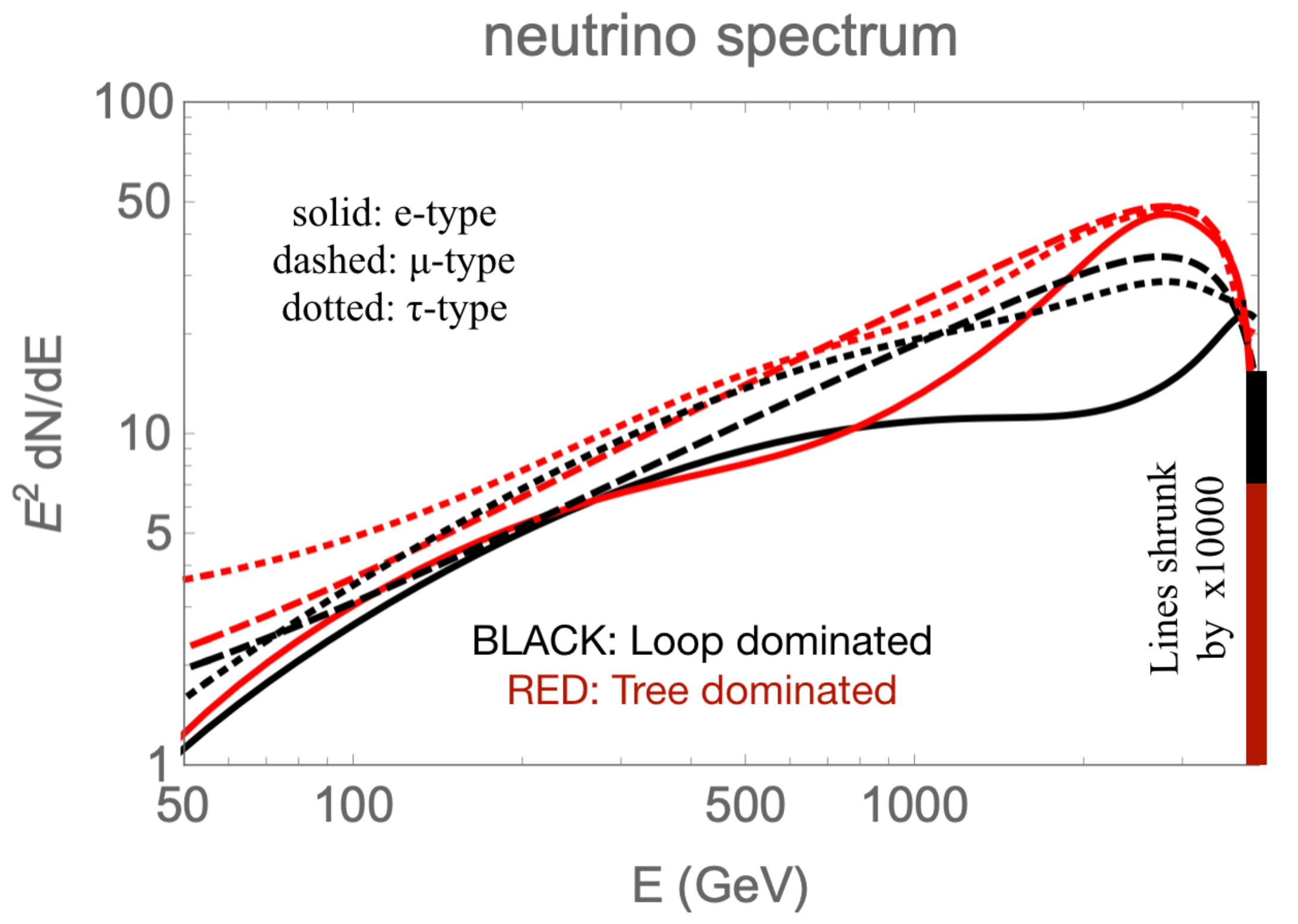}\\
\vskip0.1cm
\includegraphics[width=0.85\columnwidth]{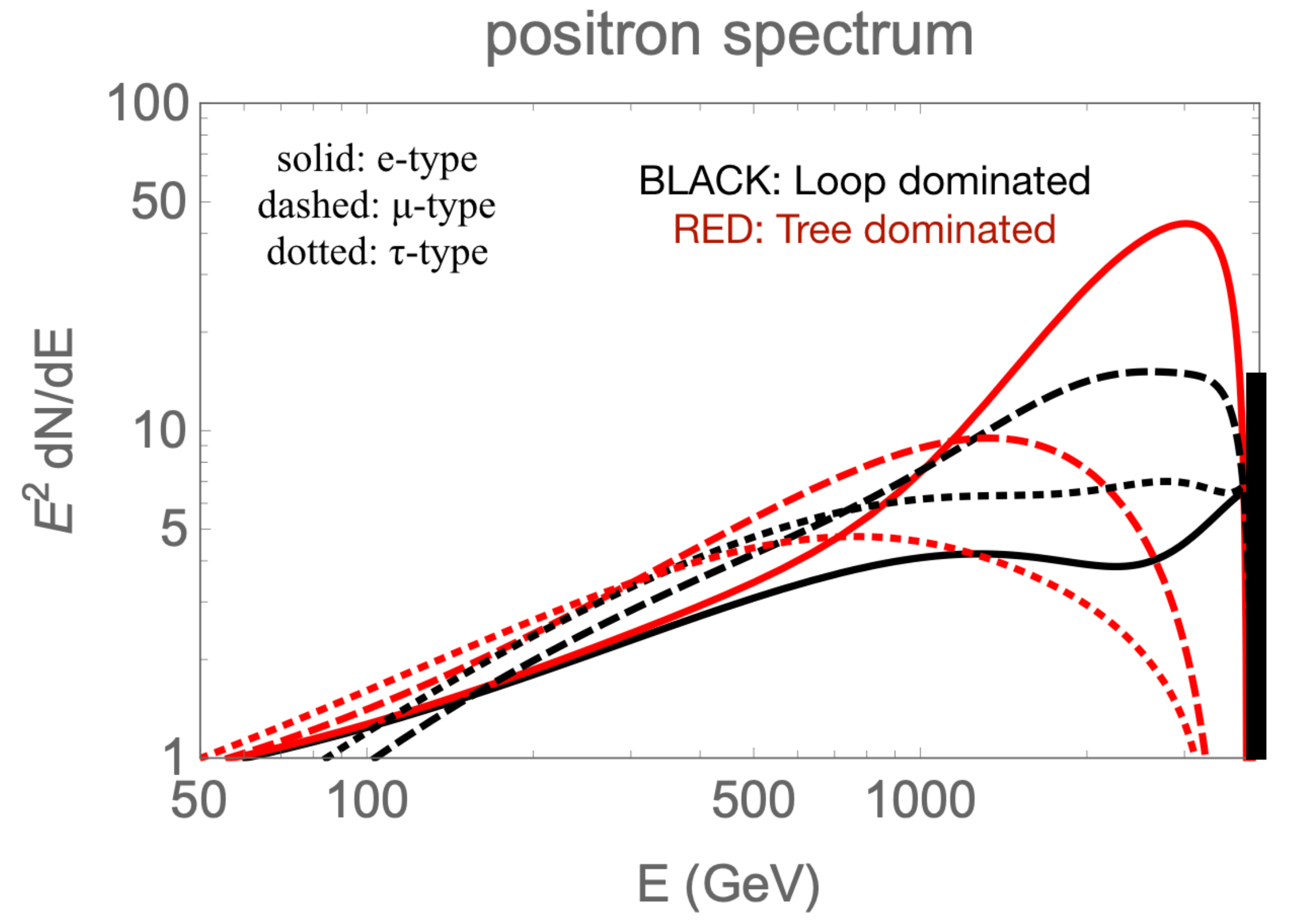}\\
\vskip0.1cm
\includegraphics[width=0.85\columnwidth]{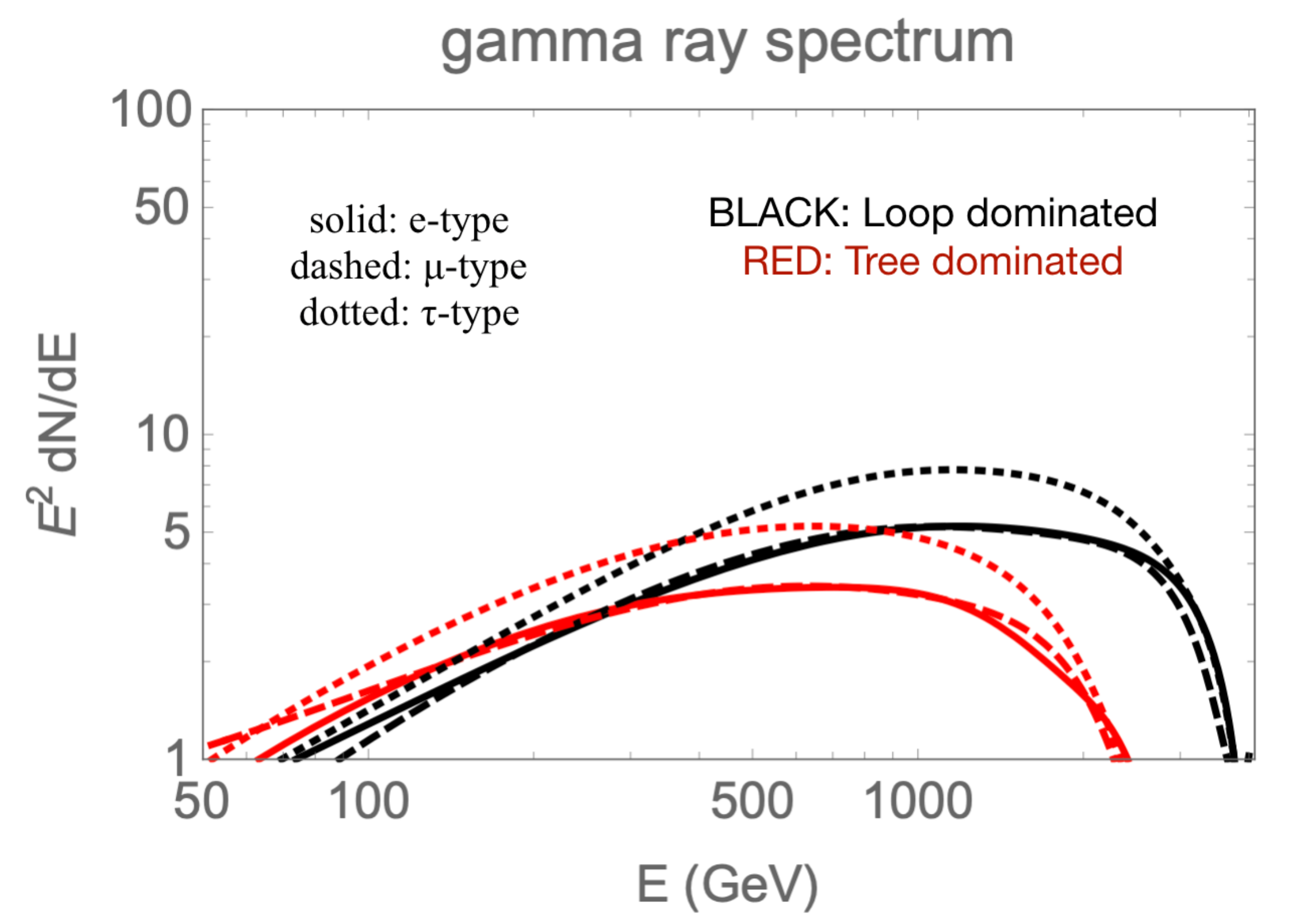}
\caption{\label{fig:spectra} Top to bottom: Spectra of neutrinos, positrons, and gamma rays produced from $m_{DM}=8$ TeV dark matter decay, in scenarios where the loop (black) or tree level (red) processes dominate, for dominant coupling to individual lepton flavors. All curves correspond to the same number of dark matter decay events, with an overall arbitrary normalization. These are prompt spectra at production as computed by Pythia \cite{Sjostrand:2006za,Sjostrand:2007gs}, and do not include propagation effects or secondary contributions such as inverse Compton scattering. Neutrino and positron line signals at $E=m_{DM}/2$ have been shrunk by a factor of $10^4$ to fit within the panels.}
\end{figure}

We now discuss indirect detection signatures for a benchmark decaying dark matter mass of $8$ TeV, for which the two-, three-, and four-body tree level decay widths are all comparable, thereby producing the most general spectrum.  As discussed earlier, due to the SU(2) nature of neutrinos, the final states also contain charged leptons as well as the SM gauge and Higgs bosons. In Figure \ref{fig:spectra}, we compare the spectra of neutrinos, positrons, and gamma rays from dark matter, assuming tree (red) or loop (black) processes are dominant, as evaluated with Pythia \cite{Sjostrand:2006za,Sjostrand:2007gs}, for dominant coupling to individual lepton flavors. We note that these are prompt spectra and do not include propagation effects (for positrons), or secondary contributions such as inverse Compton scattering or internal bremsstrahlung from within the loops (for gamma rays). All spectra correspond to the same number of events, enabling comparisons within and across the panels, but the overall normalization is arbitrary. Neutrino and positron line signals at $E=m_{DM}/2$ have been shrunk by a factor of $10^4$ to fit within the panels.  

A distinguishing feature of the loop dominated scenario is the presence of a neutrino line at $m_\text{DM}$ ($m_\text{DM}$/2) for annihilating (decaying) dark matter, which persists for arbitrarily high dark matter masses. In the plot, we also see a neutrino line signal (red) from tree level decays, as the two body decay branching fraction is still significant for this particular benchmark; however, this line would disappear for higher dark matter masses as four body decays grow to dominate. On the other hand, for tree level decays, we see hard neutrinos at approximately half to two-thirds of the energy of the neutrino line from neutrinos in three and four body decays, which are absent in the loop dominated scenario. If the coupling is dominantly to the electron-type neutrino, one also gets an analogous line in the positron spectrum in the loop dominated case; however, note that propagation effects will smear this line, making it challenging to distinguish it from the hard positron peak present in the tree level decay spectrum. 

In general, the loop processes tend to produce harder spectra of positrons and gamma rays, as seen in the plots, since all decays are into two particle final states. For both tree and loop dominated scenarios, the positron spectrum is the hardest when the sterile neutrinos dominantly couple to the electron-type neutrino, and grows progressively softer for muon-type and tau-type couplings, as can be understood from the decay channels of muons and tau leptons. 

Another salient feature of the loop dominated scenario is that the widths into neutrinos, charged leptons, and SM bosons are approximately the same, as they arise from interchanges of internal and external legs of the same loop diagram, as discussed and calculated in Section \ref{sec:calculation} (see Eqs.\,(\ref{eq:loop}),(\ref{eq:loop2})). Comparing the size of the neutrino line with the peak flux of positrons or gamma rays might therefore provide ways to distinguish between loop dominated  and tree level decays: recall that in the latter case, the ratio between two, three, and four body decay widths can be deduced from the dark matter mass (see Eq.\,\ref{eq:threebody},\ref{eq:fourbody}). This feature can also distinguish the loop dominated scenario from other frameworks not related to the heavy neutrino portal, such as, for instance, $Z'$ dark matter that couples as $Z'_\mu \bar{L}^\dag \bar\sigma^\mu L$; this model would mimic the neutrino and positron line signals, but the decays into SM bosons with comparable rates, a robust prediction of the loop dominated scenario, would be absent. 

Since the main purpose of this section is to point out the main qualitative differences in the spectra of SM particles between signals dominated by tree and loop effects, we do not delve into detailed studies of experimental sensitivity or bounds on dark matter parameters. These require additional considerations, such as inclusion of secondary emission and propagation effects, which are beyond the scope of this paper, and have been performed elsewhere, see \textit{e.g.}\ \cite{Garcia-Cely:2017oco,ElAisati:2017ppn,Campo:2017nwh,Chianese:2018ijk,Blennow:2019fhy,Dekker:2019gpe,Heeck:2019guh}.

\section{Summary}

In this paper, we considered scenarios where neutrino-related loop diagrams can dominate dark matter phenomenology despite the presence of tree-level channels. We found this feature to be fairly generic in models where the dark matter is part of a hidden sector that is connected to the SM via a heavy neutrino portal; such a portal is generic and arises in well-motivated new physics  scenarios from the point of view of hidden sectors as well as neutrino mass generation mechanisms. In such frameworks, the tree level processes, although open, are significantly suppressed by the existence of heavy sterile neutrino propagators, incurring suppression factors of powers of $v/M_N$ or $m_\text{DM}/M_N$. We demonstrated that the loop processes can overcome this suppression (provided they do not require explicit Higgs vev insertions for e.g. hypercharge conservation) and therefore dominate dark matter phenomenology in large regions of parameter space.  

While this unexpected dominance of loop processes can affect the calculation of dark matter production and lifetime, affecting compatible regions of parameter space in various models, such concerns are model-specific; more generic and interesting is the effect on dark matter indirect detection signatures, where more robust, model-independent predictions are possible. We demonstrated that the energy spectra of positrons, gamma rays, and neutrinos in loop-dominated scenarios are qualitatively different from those from tree-level decay processes. A naive calculation of dark matter signatures in these setups using only tree level processes would therefore yield extremely inaccurate predictions for the signals expected at experiments. 

We highlight two salient features of such loop dominated dark matter signatures. The first is the existence of a monochromatic neutrino line. While tree level decays also feature such lines, we found that for $m_\text{DM}\gtrsim 10\, v$, the line signal gets overwhelmed by four-body decays, which grow to dominate. Second, due to SU(2) invariance, the occurrence of analogous decays into charged leptons as well as SM gauge and Higgs bosons at comparable rates is a robust prediction of the loop-dominant scenario, which is difficult to replicate in tree-level neutrino portal dark matter models or other dark matter models that couple to lepton doublets.  The observation of monochromatic neutrino lines along with accompanying spectra in positrons or gamma rays that match the predictions of the above relations would therefore suggest that dark matter interactions are mediated by heavy sterile neutrinos, and that such loop dominated effects are dominantly at play, providing crucial insight into the underlying model.  

Given the tremendous interest in new physics related to the neutrino sector, in particular in connection with dark matter, along with the emergence of high sensitivity neutrino detectors and gamma ray experiments, we believe that it is important for the community to be aware of such unexpected but dominant effects that can occur in well motivated theoretical frameworks and offer qualitatively different yet robustly predictable dark matter indirect detection signatures that might be discovered in the coming years.

\vskip 0.2cm

\textit{Acknowledgements:} We acknowledge illuminating conversations with Wolfgang Altmannshofer. BS thanks the the GGI Institute for Theoretical Physics, and the Mainz Institute for Theoretical Physics (MITP) of the Cluster of Excellence PRISMA+ (Project ID 39083149), where parts of this research were completed, for hospitality and support. HHP and SP are partly supported by the U.S. Department of Energy grant number DE-SC0010107.

\bibliography{DM_loop}{}

\end{document}